\documentclass[12pt]{article}
\usepackage[T1]{fontenc}
\usepackage[latin1]{inputenc}
\usepackage{graphicx}
\usepackage[english]{babel}
\usepackage{amsmath}
\usepackage{amssymb}
\usepackage{amsfonts}
\title{\bf Cosmological quantum tunneling and holographic principle}
\author{F. Darabi$^1$\thanks{e-mail: f.darabi@azaruniv.edu}  and S. Jalalzadeh$^2$\thanks{email: s-jalalzadeh@sbu.ac.ir}  \\{$^1$\small Department of Physics, Azarbaijan Shahid Madani University, Tabriz 53714-161, Iran }\\ {$^1$\small Research Institute for Astronomy and Astrophysics of
Maragha (RIAAM), Maragha 55134-441, Iran.}\\{$^2$\small Department of Physics, Shahid Beheshti University, Evin, Tehran 19839,
Iran.}}
\begin{document}
\maketitle
\begin{abstract}
The probable trajectory of the ground state wave function of the universe arises through a quantum tunneling by gravitational instantons. We calculate the quantum tunneling rate for a $n>2$ dimensional closed Friedmann-Robertson-Walker universe with a positive cosmological constant. In four dimensions, by using of the holographic principle, the tunneling rate is related to the maximal entropy of the early universe after quantum tunneling.
\end{abstract}
Pacs: 98.80.Qc
\newpage
\section{Introduction}
Interest in explaining why and how the space-time has become a $4D$ manifold is not a new challenge of human mind. In fact, most of explanations made to date are essentially based on the anthropic principle \cite{anthropic}. Our existence in the universe is simply considered as an anthropic reason for four dimensionality of the space-time. If space would have more than three dimensions, no stable bound orbits of planets around the sun would
be possible and so no human kind would exist. Also in quantum mechanics the same argument applies to the existence of stable atoms which are the building blocks of every material structure.
Trying to explain the dimensionality of space-time without resorting to the
anthropic principle is an important problem which is usually supposed to be addressed within the fundamental theories like string or supersrting theory. In such fundamental theories, one assumes an $n$-dimensional space-time and by demanding some special consistency requirements in the theory the number of dimensions is obtained \cite{String}. However, a priori, it is usually assumed that four, out of n dimensions, are large ordinary space-time and the remaining ones are compact space of the Planck scale. An interesting attempt to drive dynamically a $4D$ space-time in string
theory has been made in the non-perturbative \textbf{M} formulation of type IIB strings \cite{String2}. On the other hand, there are some approaches with special interest in finding a reasonable answer and justification for the question that why four dimensional space-time has Lorentzian signature \cite{Signature}, why nature has made a choice of one time and three space coordinates \cite{space+time}, or why do we live in 3+1 dimensions \cite{3+1}.

In general relativity, our base assumption is that space-time in large scales is a $4D$ differentiable manifold. On the other hand, in small scales like
the Planck length, quantum effects will be important and space-time will be highly curved with all possible topologies and of arbitrary dimensions.
So, the main question is: how a 4-dimensional space-time is singled out of other quantum mechanically possible configurations. In the present paper, we aim to answer this question by limiting ourselves to the dimensionality,
rather than topology of space-time. To this end, we consider an $n$-dimensional empty closed\footnote{The fact that observable universe is almost spatially flat, does not mean that the topology of space corresponds exactly to $k=0$. In fact current observations can not fix the topology parameter $k$ at all. The observations just tell us that the spatial curvature $k/a^2$ is vanishing, not because of $k=0$, but because the scale factor or the radius of our universe $a$ has experienced an inflationary phase at early universe so that the spatial curvature $k/a^2$ is practically zero in the Friedmann equation. This is so called flatness problem which is solved by inflationary scenarios. In
principle, quantum tunneling for a spatially flat $k=0$ universe is impossible
because $k=0$ provides no potential barrier for tunneling. However, for a
spatially closed universe $k=1$, a potential barrier exists which leads to
quantum tunneling. After quantum tunneling from nothing to a closed FRW baby universe, the inflation provides a large scale factor in the denominator of the spatial curvature $1/a^2$ and hence the spatial curvature almost vanishes
and yields an spatially flat universe.} ($k=1$) Friedmann-Robertson-Walker (FRW) universe with a cosmological constant and then calculate the tunneling rate for creation of this universe from nothing. It turns out that the tunneling rate in four dimensional spacetime engages with the holographic principle and the maximal entropy of a FRW baby universe.

\section{The model}

We consider an $n$-dimensional empty closed ($k=1$) Friedmann-Robertson-Walker
(FRW) universe with a cosmological constant $\Lambda$. Although the origin of cosmological constant is the quantum fluctuations of vacuum, however,
it is common in cosmology to include by hand the cosmological constant. The
motivation for introducing the $\Lambda$ term is to construct a de Sitter universe. This is because in the quantum cosmology the people are interested in establishment of an expanding universe after quantum tunneling \cite{Vil}. The line element is given by
\begin{eqnarray}\label{1}
ds^2=-dt^2+a^2(t)\left[\frac{dr^2}{1-r^2}+r^2d\Omega^2_{n-2}\right],
\end{eqnarray}
where $a(t)$ as the scale factor is the only dynamical degree of
freedom and
\begin{eqnarray}\label{2}
d\Omega^2_{n-2}=d\theta^2_1+\sin^2\theta_1 d\theta^2_2 + ... +\sin^2\theta_1
\sin^2\theta_2 ... \sin^2\theta_{n-3} d\theta^2_{n-2}.
\end{eqnarray}
The spatial measure over the space-like hypersurface $\cal V$ corresponding to this metric is obtained
\begin{eqnarray}\label{3}
V&=&\int_{\cal V}\sqrt{-g}dr d\theta_1 ... d\theta_{n-3}d\theta_{n-2}\nonumber\\
&=&a^{n-1}(t)\int_0^r
\frac{r^{n-2}dr}{\sqrt{1-r^2}}\int_0^\pi (\sin\theta_1)^{n-3}d\theta_1 ...
\int_0^\pi \sin\theta_{n-3}d\theta_{n-3}\int_0^{2\pi}d\theta_{n-2}.
\end{eqnarray}
Using the following integral
\begin{eqnarray}\label{4}
\int_0^r\frac{r^{n-2}dr}{\sqrt{1-r^2}}=\frac{\sqrt{\pi}}{2}\frac{\Gamma(\frac{n-1}{2})}{\Gamma(\frac{n}{2})},
\end{eqnarray}
we obtain the measure as follows
\begin{eqnarray}\label{5}
\int_{\cal V}\sqrt{-g}dr d\theta_1 ... d\theta_{n-3}d\theta_{n-2}=a^{n-1}(t)\frac{2\pi^{\frac{n}{2}}}{\Gamma(\frac{n}{2})}.
\end{eqnarray}
The Ricci scalar corresponding to the metric (\ref{1}) is given by
\begin{eqnarray}\label{6}
{\cal R}=(n-1)\left[2\frac{\ddot{a}}{a}+(n-2)\frac{\dot{a}^2}{a^2}+(n-2)\frac{1}{a^2}\right],
\end{eqnarray}
where, for $n=4$, we recover the standard 4-dimensional Ricci scalar
\begin{eqnarray}\label{7}
{\cal R}=6\left[\frac{\ddot{a}}{a}+\frac{\dot{a}^2}{a^2}+\frac{1}{a^2}\right].
\end{eqnarray}
The gravitational action corresponding to this model is given by
\begin{eqnarray}\label{8}
S&=&\frac{1}{G_n}\int dt\: ({\cal R}-2\Lambda)\int_{\cal V}\sqrt{-g}dr d\theta_1 ... d\theta_{n-3}d\theta_{n-2}\nonumber\\
&=&M_{P}^{n-2}\frac{2\pi^{\frac{n}{2}}}{\Gamma(\frac{n}{2})}\int dt\:
a^{n-1}(t)\left[(n-1)\left(2\frac{\ddot{a}}{a}+(n-2)\frac{\dot{a}^2}{a^2}+(n-2)\frac{1}{a^2}\right)-2\Lambda\right]\nonumber,
\end{eqnarray}
where $G_n$ is the $n$-dimensional gravitational constant and, in the the units $\hbar=c=1$, the Planck mass in n-dimension is $M_{P}={G_n}^{-1/{n-2}}$. After a total derivative on the term containing $\ddot{a}$, the final
form of the action is obtained
\begin{eqnarray}\label{9}
S=-(n-1)(n-2)M_{P}^{n-2}\frac{\pi^{\frac{n}{2}}}{\Gamma(\frac{n}{2})}\int dt\: \left[a^{n-3}(\dot{a}^2-1)+\frac{2\Lambda}{(n-1)(n-2)}a^{n-1}\right].
\end{eqnarray}
Therefore, the Lagrangian is given by
\begin{eqnarray}\label{10}
{\cal L}=-\lambda \left[a^{n-3}(\dot{a}^2-1)+\frac{2\Lambda}{(n-1)(n-2)}a^{n-1}\right],
\end{eqnarray}
where
\begin{eqnarray}\label{11}
\lambda=(n-1)(n-2)M_{P}^{n-2}\frac{\pi^{\frac{n}{2}}}{\Gamma(\frac{n}{2})}.
\end{eqnarray}
The momentum conjugate to the scale factor is given by
\begin{eqnarray}\label{12}
\Pi_a=\frac{\partial {\cal L}}{\partial \dot{a}}=-2\lambda \dot{a} a^{n-3}.
\end{eqnarray}
Then, we obtain the Hamiltonian as
\begin{eqnarray}\label{13}
{\cal H}=\dot{a}\Pi_a-{\cal L}=-\frac{1}{4\lambda a^{n-3}}\pi^2_a-\lambda a^{n-3}+\frac{2\Lambda
\lambda}{(n-1)(n-2)}a^{n-1}.
\end{eqnarray}
The zero energy requirement of this gravitational system leads to the following
Hamiltonian constraint
\begin{eqnarray}\label{14}
\Pi^2_a+4\lambda^2 a_0^{2n-6}q^{2n-6}[1-q^2]=0,
\end{eqnarray}
where $q=a/a_0$ and
\begin{eqnarray}\label{15}
a_0^2=\frac{(n-1)(n-2)}{2\Lambda}.
\end{eqnarray}
Using this equation we can estimate the value of cosmological constant as
$\Lambda \sim a_0^{-2}$ which is so large considering a very small size $a_0$
for the universe after quantum tunneling. In fact, we need such a large cosmological
constant at early universe to establish a de Sitter expansion (see bellow).
Note that for $n>2$ the cosmological constant should be positive.
\section{Classical Cosmology}

Using (\ref{15}), the Lagrangian (\ref{10}) is rewritten as
\begin{eqnarray}\label{16}
{\cal L}=-\lambda \left[a^{n-3}(\dot{a}^2-1)+\frac{1}{a_0^2}a^{n-1}\right].
\end{eqnarray}
The Euler-Lagrange equation is then obtained
\begin{eqnarray}\label{17}
\frac{\ddot{a}}{a}+\frac{n-3}{2}\frac{(\dot{a}^2-1)}{a^2}-\frac{n-1}{2a_0^2}=0.
\end{eqnarray}
The Hamiltonian constraint (\ref{14}) results in the following equation
\begin{eqnarray}\label{18}
H^2+\frac{1}{a^2}-\frac{1}{a_0^2}=0,
\end{eqnarray}
where $H=\dot{a}/a$ is the Hubble parameter. Combining Eqs.(\ref{17}), (\ref{18})
we obtain
\begin{eqnarray}\label{19}
\frac{\ddot{a}}{a}-\frac{1}{a_0^2}=0.
\end{eqnarray}
The solution of this equation, subject to the Hamiltonian constraint (\ref{18}),
is obtained \cite{Vil}
\begin{eqnarray}\label{20}
a(t)=a_0 \cosh(t/a_0),
\end{eqnarray}
where $a_0$ is interpreted as the minimum radius of the universe and the initial conditions $a(0)=a_0$ and $\dot{a}(0)=0$ are used. Note that the parameter $n$ is included in solution (\ref{20}) through the definition of $a_0$ in Eq.(\ref{15}). This solution corresponds to a usual de Sitter spacetime where a phase of contraction from infinitely past time is followed by an expansion phase where the scale factor has reached its minimum $a_0$. However, a different cosmological scenario can give rise to an identical de Sitter
expansion if we consider analytic continuation $\tau=it+(\pi/2)a_0$.
Then, one can obtain the instanton solution. In fact, the Euclidean solutions of Einstein equations which are obtained by analytic continuations of the Lorentzian solutions are called instantons which is obtained here as
\begin{eqnarray}\label{21}
a_E=a_0\sin(\frac{\tau}{a_0}),
\end{eqnarray}
so that the instanton is an n-sphere of radius $a_0$ if
$\tau/a_0\in \{-\pi/2,\pi/2\}$, with spherical three-dimensional
sections labelled by the latitude angle $\theta=a_0 \tau$. Both
metrics are related by the analytic continuation into the complex
plane of the Euclidean ``time'' $\tau$
\begin{eqnarray}\label{22}
\tau=\frac{\pi}{2}a_{0}+it,\hspace{.5cm}
a=a_{E}(\frac{\pi}{2}a_0+it),
\end{eqnarray}
which is a Wick rotation with respect to the point $\tau=(\pi/2)a_0$
in this plane. This analytic continuation can be interpreted as a
quantum nucleation of the Lorentzian de Sitter spacetime from the
Euclidean hemisphere as a matching of the two manifolds across the
equatorial section $\tau=(\pi/2)a_0,\, (t=0)$, the bounce surface of
zero extrinsic curvature.

\section{Quantum Cosmology}

To calculate the quantum tunneling rate we need the "Instantons". In section 3, we obtained the Lorentzian solution (\ref{20}) and then obtained its Euclidean counterpart, namely instanton solution (\ref{21}). Now, in section 4, we calculate the tunneling rate by using the instanton solution (\ref{21}). Canonical quantization of this cosmological model in the coordinate representation is accomplished by the following operator realizations
\begin{eqnarray}\label{23}
a=a, \hspace{.5cm} \Pi_a=-i\hbar\frac{\partial}{\partial a}.
\end{eqnarray}
Then, the Hamiltonian constraint (\ref{14}) becomes the Wheeler-DeWitt equation for the wave function of the universe
\begin{eqnarray}\label{24}
\frac{d^2\psi}{dq^2}+Q^2\psi=0,
\end{eqnarray}
where $Q^2=4\lambda^2 a_0^{2(n-2)}q^{2(n-3)}[q^2-1]$, and use has been made of $q=a/a_0$ to convert the dynamical variable from $a$ to $q$.
The solutions of the Wheeler-DeWitt equation are obtained as
\begin{equation}\label{25}
\psi=\left \{ \begin{array}{ll} \frac{N}{2}|Q|^{-1/2}\exp|w| \:\:,\:\: q<1
\\
\\
{N}|Q|^{-1/2}\cos(|w|-\pi/4) \:\:,\:\: q>1
\end{array}\right.
\end{equation}
where
\begin{eqnarray}\label{26}
w=\int^q Q~dq.
\end{eqnarray}
The tunnelling rate in WKB approximation is then obtained
\begin{eqnarray}\label{27}
|T|^2=\left[1+\exp\left(2i\int_0^1 Q~ dq\right)\right]^{-1},
\end{eqnarray}
where the integral in the argument is calculated as follows
\begin{eqnarray}\label{28}
\int_0^1 Q~ dq&=&2\lambda a_0^{n-2}\int_0^1 q^{n-3}\sqrt{q^2-1}~dq\\ \nonumber
&=&i\frac{\lambda}{2}a_0^{n-2}\sqrt{\pi}\frac{\Gamma(n/2-1)}{\Gamma(n/2+1/2)}.
\end{eqnarray}
Therefore, we have
\begin{eqnarray}\label{29}
|T|^2=\left[1+\exp\left(-{\lambda}a_0^{n-2}\sqrt{\pi}\frac{\Gamma(n/2-1)}{\Gamma(n/2+1/2)}\right)\right]^{-1},
\end{eqnarray}
Substituting (\ref{11}) into (\ref{29}), and after some calculations,
we obtain the final result
\begin{eqnarray}\label{30}
|T|^2=[1+\exp(-{\alpha})]^{-1},
\end{eqnarray}
where
\begin{eqnarray}\label{31}
\alpha=\frac{4 \pi^{3/2}}{\Gamma(n/2-1/2)}[\pi a_0^2 {G}^{-1}_n]^{n/2-1}.
\end{eqnarray}
If we put $n=4$, we obtain
\begin{eqnarray}\label{32}
\alpha={8 \pi}[\pi a_0^2 {G}^{-1}_4].
\end{eqnarray}

\section{Holographic principle and tunneling rate}

Holographic principle is a property of string theory and quantum gravity which explains that the information of a volume of space can be considered
as encoded on a boundary to the region. This principle was first proposed by Gerard 't Hooft and then it was given a precise string-theory interpretation by Leonard Susskind \cite{Suss}. In a speculative sense, this theory states that the entire universe can be seen as a two-dimensional information structure "painted" on the cosmological horizon. On the other hand, the idea of holographic principle was inspired by black hole thermodynamics implying that the maximal entropy in any region scales with the radius squared. This is because the informational content of all the objects falling into the black hole can be entirely encoded in the surface fluctuations of the event horizon to resolve the black hole information paradox. In precise words, {\it the maximum information in a volume of space is proportional to the surface area of a black hole that would occupy that region}.

The multidimensional validity of holographic principle which
is relevant to this work is studied by Scardiglia and Casadiob \cite{Fabio}.
They reasoned that only in four dimensions it is possible to formulate uncertainty principles which predict the same number of degrees of freedom per spatial volume as the holographic counting, and this could be an evidence for
questioning the existence of extra dimensions. According to their claim,
when extra spatial dimensions are present, the holographic scaling is violated. However, they showed that holographic scaling can be restored if one instead allows for a violation of the equivalence principle at short distances (below the size of extra dimensions) so that the inertial mass differs from the gravitational mass in a specific non-trivial way \cite{Fabio1}. If we concern
about the exact validity of equivalence principle, we should stick to
the result that the holographic scaling is generally violated when extra spatial dimensions are present.

It is now interesting to use the holographic principle in the context of
present work. Bearing the points mentioned above, we conclude that the holographic principle, valid in four dimensions, asserts that the maximal entropy in any region scales with the radius squared of that region, namely it is proportional
to the surface area of that region. Now, we show that the term in the bracket (\ref{32}), namely $\pi a_0^2 {G}^{-1}_4$, is the maximal entropy $S$ for the early universe after quantum tunneling. To this end, we note that the surface area of the region of early universe with radius $a_0$ is $A=4\pi a_0^2$. So, Eq.(\ref{32}) can be rewritten as
\begin{eqnarray}\label{33}
\alpha={8 \pi}\frac{A}{4G},
\end{eqnarray}
or
\begin{eqnarray}\label{34}
\alpha={8 \pi}[S],
\end{eqnarray}
which leads us to 
\begin{eqnarray}\label{35}
|T|^2=[1+\exp(-{{8 \pi}[S]})]^{-1}.
\end{eqnarray}
This result indicates that the tunneling rate of a four dimensional baby
universe is related to the maximal entropy $S={A}/{4G}$ of this universe after quantum tunneling. Using (\ref{30}), we find that the tunneling rate
becomes larger for larger values of maximal entropy.
\newpage
\section{Conclusion}

By using of the gravitational instanton solution, we have calculated the quantum tunneling rate for a $n>2$-dimensional closed Friedmann-Robertson-Walker universe with a positive cosmological constant. For the special case in four dimensions and using of the holographic principle, we have obtained the interesting
result that the tunneling rate is related to the maximal entropy for the early universe after quantum tunneling. The tunneling rate becomes larger for larger values of maximal entropy. Reasonably, this indicates the tendency of quantum tunneling to occur from {\it nothing} to a FRW baby universe with {\it maximal entropy}.

\section*{Acknowledgment}
The authors would like to thank the anonymous referee for the enlightening comments. This work has been supported financially by Research Institute
for Astronomy and Astrophysics of Maragha (RIAAM) under research project
NO.1/2361.

\end{document}